\documentclass[aps,pra,twocolumn,superscriptaddress,amsmath,amssymb]{revtex4}
\usepackage{graphicx,graphics}
\usepackage{amsmath,sidecap}
\usepackage{color}
\begin{document}

\title{Gauge-potential-induced rotation of spin-orbit-coupled Bose-Einstein condensates}
\author{Jingjing Jin\footnote{jingjingjin@nuc.edu.cn}}
\affiliation{School of Science, North University of China, Taiyuan 030051, China}
\author{Wei Han\footnote{hanwei.irain@gmail.com}}
\affiliation{Key Laboratory of Time and Frequency Primary Standards, National Time Service Center, Chinese Academy of Sciences, Xi'an 710600, China}
\author{Suying Zhang}
\affiliation{Institute of Theoretical Physics, Shanxi University, Taiyuan 030006, China}

\begin{abstract}
We demonstrate that a spin-orbit-coupled Bose-Einstein condensate can be effectively rotated by adding a real magnetic field to inputting gauge angular momentum, which is distinctly different from the traditional ways of rotation by stirring or Raman laser dressing to inputting canonical angular momentum. The gauge angular momentum is accompanied by the spontaneous generation of equal and opposite canonical angular momentum in the ground states, and it leads to the nucleation of quantized vortices. We explain this by indicating that the effective rotation with the vortex nucleation results from the effective magnetic flux induced by the gauge potential, which is essentially different from the previous scheme of creating vortices by synthetic magnetic fields. In the weakly interacting regime, symmetrically placed domains separated by vortex lines as well as half-integer giant vortices are discovered. With relatively strong interatomic interaction, we predict a structure of coaxially arranged annular vortex arrays, which is in stark contrast to the familiar Abrikosov vortex lattice. The developed way of rotation may be extended to a more general gauge system.
\end{abstract}

\maketitle

\section{Introduction}
The rotation of degenerate quantum gases has always been a subject of intense interest in cold atom physics~\cite{NRCooper,ALFetter}. It plays an important role in the research of quantized vortices and superfluidity~\cite{JDalibard,WKetterle,WKetterle2,JWPan}, skyrmions and magnetism~\cite{EACornell,MUeda}, as well as quantum Hall physics~\cite{NKWilkin,TLHo,PZoller,EACornell2}. All of these aspects have significant overlap with the hot topics of superfluids~\cite{RJDonnelly} and superconductors~\cite{VMVinokur}, magnetic materials~\cite{MRho,YTokura}, as well as condensed matter physics~\cite{SMGirvin,APinczuk}. In physics, rotation is characterized by nonzero angular momentum. The usual ways of rotating a degenerate gas, involving stirring~\cite{JDalibard,WKetterle} and Raman laser dressing~\cite{IBSpielman}, are designed from the point of inputting canonical angular momentum by adding a term with $H^{\mathrm{(c)}}\sim\int\mathbf{\Psi }^{\dag }(\mathbf{r}\times \hat{\mathbf{p}})_{z}\mathbf{\Psi }d\mathbf{r}$ in the Hamiltonian.

The recent experimental realization of spin-orbit (SO) coupling in ultra-cold atomic gases~\cite{IBSpielman2,Jing-Zhang,MWZwierlein,Jing-Zhang2,Shuai-Chen,Jun-Ye,WKetterle6} has stimulated much theoretical and experimental activity~\cite{VGalitski,JDalibard2,NGoldman,Hui-Zhai,Hui-Zhai2,WYi,Jing-Zhang3,Congjun-Wu}. Considering the inherent complexity of the SO-coupled system, the traditional methods of rotation face many difficulties and challenges. For example, if we wanted to get a time-independent Hamiltonian, stirring a SO-coupled condensate would involve rotating not only the trap but also the Raman laser beams (and perhaps also the external magnetic field)~\cite{VGalitski2,ALFetter2}. Therefore, it is desirable to develop other experimentally relevant methods to rotate an ultra-cold atomic system with SO coupling.

As a gauge system, the rotation properties of a SO-coupled condensate is governed not only by the canonical angular momentum $\hat{L}_{z}^{\mathrm{c}}=(\mathbf{r}\times \hat{\mathbf{p}})_{z}$, but also by the gauge angular momentum $\hat{L}_{z}^{\mathrm{g}}=-(\mathbf{r}\times \mathbf{A})_{z}$ with $\mathbf{A}$ being the corresponding gauge potential. Both constitute the total mechanical angular momentum $\hat{L}_{z}^{\mathrm{mech}}=\hat{L}_{z}^{\mathrm{c}}+\hat{L}_{z}^{\mathrm{g}}$. This inspires a belief that the system can also gain nonzero gauge angular momentum when a term $H^{\mathrm{(g)}}\sim\int\mathbf{\Psi }^{\dag }(\mathbf{r}\times \mathbf{A})_{z}\mathbf{\Psi }d\mathbf{r}$ is added in the Hamiltonian. Previously, while the rotation properties of a SO-coupled condensate under $H^{\mathrm{(c)}}$ were examined~\cite{JHHan}, the combined effects of $H^{\mathrm{(c)}}$ and $H^{\mathrm{(g)}}$ have been also discussed~\cite{Congjun-Wu2}. It is worth noting that a pure gauge-rotated SO-coupled system (i.e., the Hamiltonian with only $H^{\mathrm{(g)}}$, which can be easily realized by a gradient magnetic field or spatially dependent Rabi coupling~\cite{IBSpielman3,Congjun-Wu2,VGalitski2,SStringari2}) is more interesting, which will not only circumvent the difficulties and challenges faced by the canonical rotation, but may also bring new quantum hydrodynamics. It has been shown that nonzero gauge angular momentum gives rise to the violation of the irrotationality constraint of superfluid velocity field~\cite{SStringari} and the precession of dipole oscillation~\cite{SStringari2}.

In this article, we demonstrate that a Dresselhaus SO-coupled condensate can be effectively rotated by an additional Ioffe-Pritchard (IP) magnetic field, whose coupling with spin equivalently inputs gauge angular momentum in the system. It is also found that the gauge angular momentum is accompanied by spontaneous generation of equal and opposite canonical angular momentum in the ground states and induces exotic static vortex arrangement. In the weakly interacting regime, symmetrically placed domains separated by vortex lines as well as half-integer giant vortices are discovered, which are very similar to those obtained in a traditional way of directly inputting canonical angular momentum~\cite{JHHan}. With relatively strong interatomic interaction, the vortices prefer to arrange themselves in concentric circles, in contrast to the familiar Abrikosov vortex lattice predicted in superconductors~\cite{AAAbrikosov} and conventional superfluids~\cite{WKetterle}. We explain the rotational effect with the nucleation of quantized vortices by indicating that the gauge potential around a closed loop accumulates a Aharonov-Bohm geometric phase factor~\cite{YAharonov,IBSpielman4} in the wave function, and induces nonzero effective magnetic flux inside the loop. This is essentially different from the previous scheme of creating vortices by synthetic magnetic fields~\cite{IBSpielman}.

The rest of the paper is organized as follows. In Sec.~\uppercase\expandafter{\romannumeral2}, we introduce the model Hamiltonian of the Dresselhaus SO-coupled Bose-Einstein condensates (BECs) in an IP magnetic field. In Sec. \uppercase\expandafter{\romannumeral3}, by analyzing the relationship between the external magnetic field, the spin vector, the canonical and gauge particle currents, as well as the effective gauge potential, we reveal the mechanism of the effective rotation that giving rise to the nucleation of quantized vortices. In Sec. \uppercase\expandafter{\romannumeral4}, based on numerical simulations, we present possible exotic many-body quantum phases, involving the symmetrically placed domains separated by vortex lines, the half-integer giant vortices and the coaxially arranged annular vortex arrays. The experimental feasibility is discussed in Sec. \uppercase\expandafter{\romannumeral5}. Finally, we summarize and give concluding remarks in Sec. \uppercase\expandafter{\romannumeral6}.

\section{Model}
We consider two-dimensional (2D) Dresselhaus SO-coupled BECs in a IP magnetic field~\cite{MSIoffe,DEPritchard}. The Hamiltonian in the Gross-Pitaevskii mean-field approximation can be written as
\begin{eqnarray}
&&H=\int\!\! d\mathbf{r}\mathbf{\Psi }^{\dag }\!\left(\!-\frac{\hbar ^{2}%
\boldsymbol{\nabla }^{2}}{2M}+V+\mathcal{V}_{\text{so}}+\mu_{B}g_{F}\boldsymbol{\sigma}\cdot \mathbf{B}\!\right)\!\mathbf{\Psi }\notag \\
&&+\frac{1}{2}\!\int\!\! d\mathbf{r}\sum\limits_{i,j=\uparrow ,\downarrow
}g_{ij}\!\Psi _{i}^{\ast }\!\left( \mathbf{r}\right)\!\Psi
_{j}^{\ast }\!\left( \mathbf{r}\right)\!\Psi _{j}\!\left( \mathbf{r}%
\right)\!\Psi _{i}\!\left( \mathbf{r}\right),\quad\label{Model Hamiltonian}
\end{eqnarray}
where $\mathbf{\Psi}=[\Psi_{\uparrow}(\mathbf{r}),\Psi_{\downarrow}(\mathbf{r})]^\top$ with $\mathbf{r}=(x,y)$ denotes the spinor order parameter, and is normalized to satisfy $\int d\mathbf{r} \mathbf{\Psi}^{\dag}\mathbf{\Psi}=~N$. The harmonic potential for trapping the atoms is $V=\frac{1}{2}M\omega_{\perp}^2(x^2+y^2)$, with $M$ the atomic mass and $\omega_{\perp}$ the trapping frequency. The Dresselhaus SO coupling term is written as
$\mathcal{V}_{\text{so}}=-i\hbar\kappa(\sigma_{x}\partial_{y}+\sigma_{y}\partial_{x})$~\cite{Hui-Zhai},
where $\sigma_{x,y}$ are the components of the Pauli matrix vector $\boldsymbol{\sigma}$ and $\kappa$ denotes the SO coupling strength. The IP magnetic field is often expressed as $\mathbf{B}=B'\left( x\hat{x}-y\hat{y}\right)+B_{z}\hat{z}$, where $B'$ is the magnetic field gradient in the 2D plane, and we focus on the case of axial bias field $B_{z}=0$. The coupling between the magnetic field and spin is related to the Bohr magneton $\mu_{B}$ and the Land\'{e} factor $g_{F}$.

In real BEC experiments, the IP magnetic field has been successfully used in trapping atoms~\cite{WKetterle3} and dynamically imprinting vortices~\cite{WKetterle4} and spin textures~\cite{WKetterle5,MMottonen}. The Dresselhaus SO coupling may be experimentally created by Raman laser dressing~\cite{GJuzeliunas,IBSpielman2,ZFXu} or modulating gradient magnetic field~\cite{LYou,GJuzeliunas2,Ruquan-Wang}, and has been recently realized in $^{40}$K degenerate Fermi gases~\cite{Jing-Zhang2}. The 2D geometry can be realized by imposing a strong harmonic potential $V(z)=M\omega_{z}^2z^2/2$ along the axial direction with $\omega_{z}\gg \omega_{\perp}$, in which case the effective contact-interaction strength is given by $g_{ij}=\sqrt{8\pi}(\hbar^2/M)(a_{ij}/a_{h_{z}})$ with $a_{ij}$ being the $s$-wave scattering length and $a_{h_z}=\sqrt{\hbar/M\omega_{z}}$ the axial characteristic length~\cite{JDalibard3}.

\section{Gauge-potential-induced rotation}
\begin{figure}[tbp]
\centerline{\includegraphics[width=0.45\textwidth,clip=]{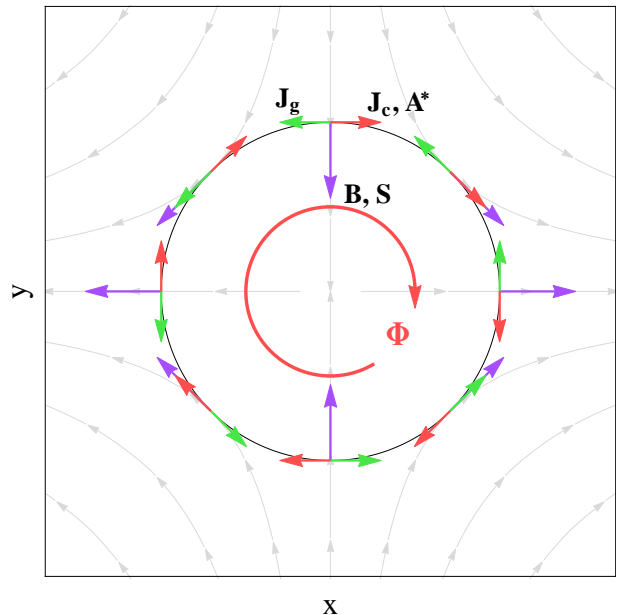}}
\caption{Scheme of Ioffe-Pritchard magnetic-field-induced rotation in a system with Dresselhaus spin-orbit coupling. The gray lines and gray arrows depict the direction of the magnetic field in the $x$-$y$ plane. Arrows with other colors depict the direction of the magnetic field $\mathbf{B}$ and spin $\mathbf{S}$ (pruple), gauge angular momentum $\mathbf{J}_{\mathrm{g}}$ (green), canonical angular momentum $\mathbf{J}_{\mathrm{c}}$ and effective gauge potential $\mathbf{A}^{*}$ (red), respectively. The integral of $\mathbf{A}^{*}$ around a closed loop induces a Aharonov-Bohm geometric phase factor $\exp{\left(i\Phi/\hbar\right)}$ in the wave function, where the effective magnetic flux $\Phi=\oint\mathbf{A}^{*}\cdot d\mathbf{l}$ is responsible for the generation of quantized vortices.}\label{fig1}
\end{figure}

For atoms with a negative Land\'{e} $g$ factor, it is energy favored for the spin vector $\mathbf{S}=\mathbf{\Psi }^{\dag }\boldsymbol{\sigma}\mathbf{\Psi }/|\mathbf{\Psi }|^2$ of the condensate being parallel to the local magnetic field $\mathbf{B}$. At the same time, the gauge part of the particle current depends on the gauge potential $\mathbf{A}$ and is defined as $\mathbf{J}_{\mathrm{g}}=-\frac{1}{M}\mathbf{\Psi }^{\dag }\mathbf{A}\mathbf{\Psi }$~\cite{JHHan2,GBaym,ALFetter2}. For the Dresselhaus SO coupling, the gauge potential $\mathbf{A}=-\kappa M(\sigma_{y},\sigma_{x})$, and thus we have $\mathbf{J}_{\mathrm{g}}=\kappa\rho(S_{y},S_{x})$, where $\rho$ denotes the total density. As the spin is polarized in the $x$-$y$ plane by the IP magnetic field with $B_{z}=0$, a circulating gauge particle current with $\mathbf{J}_{\mathrm{g}}=\kappa \rho\hat{e}_{\varphi}$ will be induced as shown in Fig.~\ref{fig1}.

According to the hydrodynamic theory, the contribution of the mechanical movement to the Hamiltonian can be written as $H^{\mathrm{mech}}=\int M(\mathbf{J}_{\mathrm{c}}+\mathbf{J}_{\mathrm{g}})^2/2\rho d\mathbf{r}$~\cite{JHHan2}, where $\mathbf{J}_{\mathrm{c}}=\rho \mathbf{v}$ represents the canonical particle current with $\mathbf{v}=(\rho_{\uparrow}\frac{\hbar}{M}\boldsymbol{\nabla}\theta_{\uparrow}+\rho_{\downarrow}\frac{\hbar}{M}\boldsymbol{\nabla}\theta_{\downarrow})/\rho$ being the superflow velocity. Here, $\rho_{\uparrow,\downarrow}$ and $\theta_{\uparrow,\downarrow}$ denote the density and phase of each component, respectively. The energy minimization in the ground states requires that the total particle current $\mathbf{J}^{\mathrm{mech}}=\mathbf{J}_{\mathrm{c}}+\mathbf{J}_{\mathrm{g}}=0$~\cite{JHHan2}. This suggests that the emergence of circulating gauge particle current is accompanied by an equivalent canonical particle current in its opposite direction, as shown in Fig.~\ref{fig1}. While the gauge particle current $\mathbf{J}_{\mathrm{g}}$ depends on the spin, the canonical part $\mathbf{J}_{\mathrm{c}}$ is related to the phase gradient, and thus may be sustained by the generation of quantized vortices.

In the Dresselhaus spin-orbit coupled system, the effective value of the gauge potential $\mathbf{A}^{*}=\mathbf{\Psi }^{\dag }\mathbf{A}\mathbf{\Psi }/|\mathbf{\Psi }|^2$ depends on the spin vector $\mathbf{S}$ with $\mathbf{A}^{*}=-\kappa M(S_{y}, S_{x})$. As the spin is polarized in the $x$-$y$ plane, as shown in Fig.~\ref{fig1}, we have $\mathbf{A}^{*}=-\kappa M \hat{e}_{\varphi}$. As a result, the integral of $\mathbf{A}^{*}$ around a closed loop with radius $R$ will lead to a Aharonov-Bohm geometric phase factor $\exp{\left(i\Phi/\hbar\right)}$ in the wave function, with
\begin{equation}
\Phi=\oint\mathbf{A}^{*}\cdot d\mathbf{l}=M\oint\mathbf{v}\cdot d\mathbf{l}=-2\pi R\kappa M\label{effective magnetic flux}
\end{equation}
being an effective magnetic flux. In the next section, we will illuminate that this effective magnetic flux will give rise to novel vortex phases in the many-body ground states, where the total vortex quantum number can be estimated by $N_{\mathrm{v}}=\Phi/h$.
\begin{figure}[t]
\centerline{\includegraphics[width=0.48\textwidth,clip=]{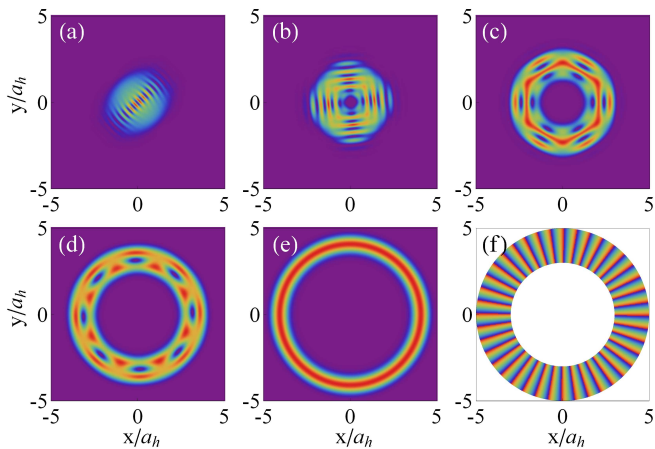}}
\caption{Ground state as a function of magnetic field gradient. (a)-(d) Density distributions of the symmetrically placed domains separated by vortex lines with (a) $B^{\prime }=0.2\ \hbar^{2}/\mu_{B}g_{F}Ma_{h}^3$, (b) $B^{\prime }=1.2\ \hbar^{2}/\mu_{B}g_{F}Ma_{h}^3$, (c) $B^{\prime }=2.2\ \hbar^{2}/\mu_{B}g_{F}Ma_{h}^3$, and (d) $B^{\prime }=3.2\ \hbar^{2}/\mu_{B}g_{F}Ma_{h}^3$. (e)-(f) Density and phase distributions of the half-integer giant vortex with $B^{\prime}=4\ \hbar^{2}/\mu_{B}g_{F}Ma_{h}^3$. Other parameters are fixed at $\kappa=10\ \hbar/Ma_{h}$, $Ng_{\uparrow\uparrow }=Ng_{\downarrow\downarrow }=10\ \hbar^{2}/M$ and $Ng_{\uparrow \downarrow }=8\ \hbar^{2}/M$. Here, $a_{h}=\sqrt{\hbar/M\omega_{\perp}}$ is the characteristic length of the harmonic trap.}
\label{fig2}
\end{figure}

The mechanism of the effective rotation can be understood by noting that the magnetic-field-spin-coupling term $H^{\textrm{M-S}}=\mu_{B}g_{F}\int d\mathbf{r}\mathbf{\Psi }^{\dag }(\boldsymbol{\sigma}\cdot \mathbf{B})\mathbf{\Psi }$ in the Hamiltonian equivalently inputs gauge angular momentum $L_{z}^{\mathrm{g}}=-\int\mathbf{\Psi }^{\dag }(\mathbf{r}\times\mathbf{A})_{z}\mathbf{\Psi }d\mathbf{r}$ in the system with Dresselhaus SO coupling. This causes rotation effects of the condensates, and also induces equal and opposite canonical angular momentum in the ground states. This is completely different from the traditional manner of rotation of a condensate by adding a term $H^{\mathrm{(g)}}\sim\int\mathbf{\Psi }^{\dag }(\mathbf{r}\times \mathbf{A})_{z}\mathbf{\Psi }d\mathbf{r}$ in the Hamiltonian to inputting canonical angular momentum $L_{z}^{\mathrm{c}}=\int\mathbf{\Psi }^{\dag }(\mathbf{r}\times\mathbf{p})_{z}\mathbf{\Psi }d\mathbf{r}$ in the system~\cite{JDalibard,WKetterle,IBSpielman}.

\section{Many-body ground states}
We next investigate the many-body ground sates of SO-coupled BECs under the gauge-potential-induced rotation, which can be calculated by numerically minimizing the Hamiltonian functional given by Eq. (1). In the weakly interacting regime, it is found that the condensates are divided into several symmetrically place domains, with radial vortex arrays playing the role of domain walls. The domain number increases with increasing the magnetic field gradient $B'$, as shown in Figs.~\ref{fig2}(a)-\ref{fig2}(d). The vortices in each component have winding number $n_{\mathrm{w}}=1$, except that in the center hole, which has winding number $n_{\mathrm{w}}> 1$ and forms a multi-quantum vortex. This structure is very similar to those obtained in systems under canonical rotation~\cite{JHHan,Congjun-Wu2}, but where the number of domains is dominated by the rotating angular frequency.
\begin{figure}[t]
\centerline{\includegraphics[width=0.48\textwidth,clip=]{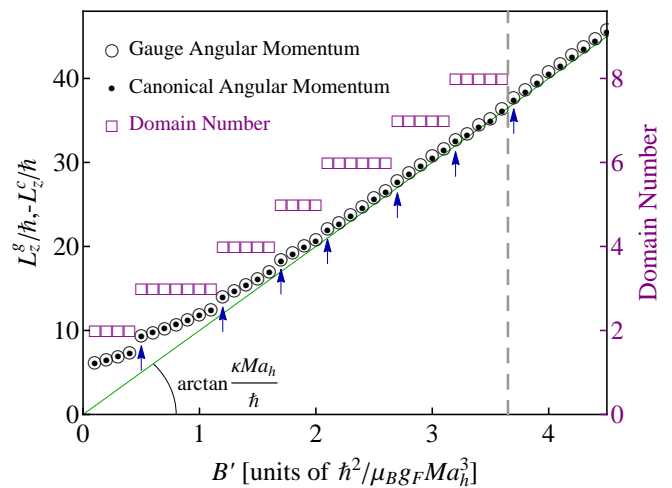}}
\caption{Gauge angular momentum $L_{z}^{\mathrm{g}}$ and canonical angular momentum $L_{z}^{\mathrm{c}}$ per particle as well as domain number as functions of the field
gradient $B^{\prime }$. Arrows indicate the jump behavior of the angular
momenta in which the domain number changes. Solid line plots the proportional function of $B'$ with slope $\kappa Ma_{h}/\hbar$. Dashed line distinguishes the regions of the domain structure and the half-integer giant vortex. Other parameters are fixed at $\kappa=10\ \hbar/Ma_{h}$, $Ng_{\uparrow\uparrow }=Ng_{\downarrow\downarrow }=10\ \hbar^{2}/M$, and $Ng_{\uparrow \downarrow }=8\ \hbar^{2}/M$.}
\label{fig3}
\end{figure}

As the magnetic field gradient increases, all the vortices gather in the center hole, forming a giant vortex structure, as shown in Figs.~\ref{fig2}(e)-\ref{fig2}(f), which has been previously observed in SO systems with external rotation~\cite{JHHan,Congjun-Wu2} or a toroidal trap~\cite{XFZhang,Yongping-Zhang}. Owing to the presence of SO coupling and IP magnetic field, the winding numbers of the giant vortices of the spin-up and -down components always differ by $1$. This can be explained by representing the wave functions as density and phase $\Psi_{j}=\sqrt{\rho_{j}}\exp(i\theta_{j})$ in the polar coordinate $(r, \varphi)$ representation. The Hamiltonian related to the relative phase involves the SO coupling and IP magnetic field terms, and can be rewritten as
\begin{eqnarray}
H^{\mathrm{RP}}\!=\!-2\kappa\!\!\int\!\!\!d\mathbf{r}\!\!\left[\sqrt{\rho_{\uparrow}}\frac{\partial\sqrt{\rho_{\downarrow}}}{\partial r}\!-\!\sqrt{\rho_{\uparrow}\rho_{\downarrow}}\frac{\partial\theta_{\downarrow}}{r\partial\varphi}\right]\!\!\cos(\theta_{\uparrow}\!-\!\theta_{\downarrow}\!-\!\varphi)\notag \\
+2B'\mu_{B}g_{F}\!\!\int\!\!\!d\mathbf{r} \sqrt{\rho_{\uparrow}\rho_{\downarrow}}r\cos(\theta_{\uparrow}\!-\!\theta_{\downarrow}\!-\!\varphi).\label{SOC Hamiltonian 2}
\end{eqnarray}
In order to satisfying energy minimization, it is required that
\begin{eqnarray}
\theta_{\uparrow}-\theta_{\downarrow}-\varphi=2\pi l, (l\in Z)\label{relative phase}
\end{eqnarray}
with $\partial\theta_{j}/\partial\varphi<0$. Thus, the giant vortex can be represented as $\mathbf{\Psi }=[\sqrt{\rho_{\uparrow}}e^{
-im\varphi},\sqrt{\rho_{\downarrow}}e^{-i(m+1)\varphi}]^\top$ with $m\in Z^{+}$.
\begin{figure}[t]
\centerline{\includegraphics[width=0.48\textwidth,clip=]{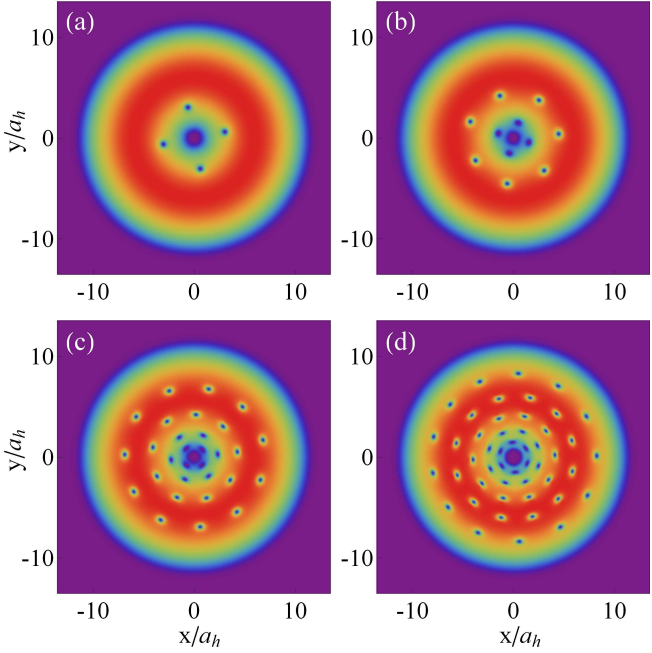}}
\caption{Ground-state density distribution as a function of the spin-orbit coupling strength. (a)-(d) Coaxially arranged vortex arrays with (a) $\kappa=1\ \hbar/Ma_{h}$, (b) $\kappa=2\ \hbar/Ma_{h}$, (c) $\kappa=4\ \hbar/Ma_{h}$, and (d) $\kappa=6\ \hbar/Ma_{h}$. Other parameters are fixed at $B^{\prime }=6\ \hbar^{2}/\mu_{B}g_{F}Ma_{h}^3$, $Ng_{\uparrow\uparrow }=Ng_{\downarrow\downarrow }=5000\ \hbar^{2}/M$, and $Ng_{\uparrow \downarrow }=4000\ \hbar^{2}/M$.}
\label{fig4}
\end{figure}

The IP magnetic field with $B_{z}=0$ polarizes the spin in the $x$-$y$ plane with $S_{z}=0$, so we have $\rho_{\uparrow}=\rho_{\downarrow}$ and $\mathbf{v}=\frac{\hbar}{2M}\boldsymbol{\nabla}\left(\theta_{\uparrow}+\theta_{\downarrow}\right)$. The circulation of the superfluid velocity along a closed path is $\oint \mathbf{v}\cdot d\mathbf{l}=\Phi/M=-(m+\frac{1}{2})h/M$. This implies that the state in Figs.~\ref{fig2}(e)-\ref{fig2}(f) is essentially a half-integer giant vortex and behaves as a physical effect of the magnetic flux $\Phi$ induced by the gauge potential $\mathbf{A}^{*}$. According to the minimum of the real-space potential energy $E(r)=\frac{1}{2}M\omega_{\perp}^2r^2\pm \mu_{B}g_{F}B'r$ caused by the harmonic and IP magnetic traps, one can estimate the radius of the giant vortex as $r_{0}=-\frac{\mu_{B}g_{F}B'}{M\omega_{\perp}^2}$. As a result, the corresponding gauge angular momentum can be expressed as $L_{z}^{\mathrm{g}}=\kappa M\int \rho r d\mathbf{r}\approx -\frac{N\mu_{B}g_{F}\kappa B'}{\omega_{\perp}^2}$, which induces equivalent canonical angular momentum $L_{z}^{c}=N(m+\frac{1}{2})\hbar$ with vortex winding number $m=-\frac{\mu_{B}g_{F}\kappa B'}{\hbar\omega_{\perp}^2}-\frac{1}{2}$.

Numerical results of the gauge and canonical angular momenta as functions of the magnetic field gradient $B'$ are shown in Fig.~\ref{fig3}. It is found that both the gauge and canonical momenta are asymptotically proportional to the SO coupling strength $\kappa$ and the magnetic field gradient $B'$, which is consistent with the above analytical analysis. In addition, we also observe slight jumps of the angular momentum (indicated by arrows in Fig.~\ref{fig3}) where the domain number changes~\cite{JHHan}.
\begin{figure}[t]
\centerline{\includegraphics[width=0.48\textwidth,clip=]{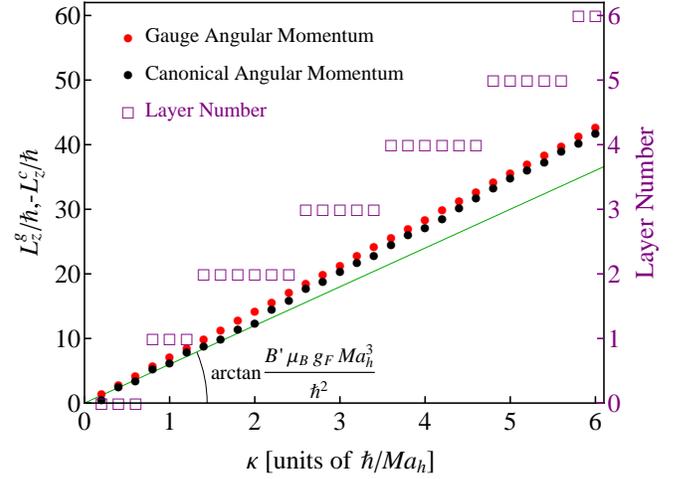}}
\caption{Gauge angular momentum $L_{z}^{\mathrm{g}}$ and
canonical angular momentum $L_{z}^{\mathrm{c}}$ per particle as well as layer number as functions of the spin-orbit coupling strength $\kappa$. Green solid line plots the proportional function of $\kappa$ with slope $B'\mu_{B}g_{F}Ma_{h}^3/\hbar^2$. Other parameters are fixed at $B^{\prime }=6\ \hbar^{2}/\mu_{B}g_{F}Ma_{h}^3$, $Ng_{\uparrow\uparrow }=Ng_{\downarrow\downarrow }=5000\ \hbar^{2}/M$, and $Ng_{\uparrow \downarrow }=4000\ \hbar^{2}/M$.}
\label{fig5}
\end{figure}

In the parameter region with relatively strong interatomic interactions, the ground-state vortex arrangement caused by the effective magnetic flux $\Phi$ is very peculiar. It is found that the vortices prefer to arrange themselves as coaxially arranged annular arrays as shown in Fig.~\ref{fig4}. The layer number of the annular vortex arrays increases with the SO coupling strength $\kappa$. This is not only different from the Abrikosov triangular lattice of superconductors~\cite{AAAbrikosov}, but also from the square or hexagonal lattice discovered in traditional multi-component superfluids~\cite{EACornell,TLHo2,MUeda2,NRCooper2,Hui-Zhai3}. It should also be emphasized that in the coaxially annular arrays all the vortices take the same direction of circulation which different from those spontaneous vortex lattices in a irrotational SO-coupled system, where vortices and antivortices emerge in pairs~\cite{LSantos,Hui-Hu,LYou2,SCGou,MUeda3,Wei-Han}.

Accompanying the vortex nucleation, equal and opposite gauge and canonical angular momenta are generated whose absolute values are approximately proportional to the SO coupling strength, with slopes greater than the magnetic field gradient $B'$, as shown in Fig.~\ref{fig5}. The inconsistency of the slopes results from the influence of the interatomic interaction, which enlarges the radius of gyration of the condensate.

\section{Experimental relevance}
The experimental realization of the model Hamiltonian in Eq.~(\ref{Model Hamiltonian}) may be achieved with the (5S$_{1/2}$, $F=1$) ground electronic manifold of $^{87}$Rb atoms, where two of the hyperfine states, $|F=1,m_{F}=-1>$ and $|F=1,m_{F}=0>$, are chosen to simulate the spin-up $|\uparrow>$ and spin-down $|\downarrow>$ components, respectively. The corresponding Land\'{e} factor is $g_{F}=-1/2$. The Dresselhaus spin-orbit coupling can be induced by Raman laser pulses or magnetic-field-gradient pulses as suggested in Refs.~\cite{ZFXu,LYou,GJuzeliunas2}. In the schemes, the spin-orbit coupling strength is determined from the Raman laser wavelength or the magnetic field gradient with the attainable value of approximately $\kappa M/\hbar\approx 1\ \mathrm{\mu m}^{-1}$. To observe the predicted vortex structures, such as the giant vortex with winding number $m$, it is required that $\kappa B'=-m\hbar\omega_{\perp}^2/\mu_{B}g_{F}$, which is readily attainable with current technologies by adjusting the magnetic field gradient $B'=m\times 10^{-4}\mathrm{G}/\mathrm{\mu m}$ and taking the trapping frequency $\omega_{\perp}\sim 2\pi\times 10^2\ \mathrm{Hz}$~\cite{WKetterle4}.

\section{Conclusion}
In summary, we have developed a new method to rotate systems with spin-orbit coupling. It is suggested that a gauge system can be effectively rotated by inputting gauge angular momentum $\hat{L}_{z}^{\mathrm{g}}=-(\mathbf{r}\times \hat{\mathbf{A}})_{z}$ instead of inputting canonical angular momentum $\hat{L}_{z}^{\mathrm{c}}=(\mathbf{r}\times \hat{\mathbf{p}})_{z}$. In particular, we investigate the gauge-potential-induced rotation of Dresselhaus spin-orbit-coupled Bose-Einstein condensates, where the gauge angular momentum is input by an additional Ioffe-Pritchard magnetic field. The many-body ground states are discussed, and it is found that equal and opposite canonical angular momentum is induced for energy minimization of the mechanical movement and accompanied
by an exotic vortex arrangement that is different from the usual Abrikosov vortex lattice. It should be emphasized that the effective rotation with the nucleation of quantized vortices results from the effective magnetic flux induced by the gauge potential, which is essentially different from the previous scheme of creating vortices by synthetic magnetic fields~\cite{IBSpielman}. The developed method circumvents the difficulties and challenges~\cite{VGalitski2,ALFetter2} faced by the traditional ways of rotation for a spin-orbit-coupled system, and brings new perspectives on the physics of ultra-cold atomic gases under gauge potentials.

Finally, the present investigation may be generalized to systems with general gauge potentials (including, but not limited to, spin-orbit coupling), where a possible scheme of gauge-potential-induced rotation may be realized by adding a term with $H^{\mathrm{add}}\sim\int\mathbf{\Psi }^{\dag }(\mathbf{r}\times\mathbf{A})_{z}\mathbf{\Psi }d\mathbf{r}$ in the Hamiltonian to inputting non-zero gauge angular momentum. In addition, we note the vortex nucleation of condensates with other types of spin-orbit coupling induced by gradient magnetic field or spatially varying laser detuning~\cite{IBSpielman3,VGalitski2,Congjun-Wu3}. These, in fact, can also be understood by noting that the gradient magnetic field or spatially varying laser detuning inputs gauge angular momentum in those systems.

\section*{ACKNOWLEDGMENTS}
This work was supported by the National Natural Science Foundation of China under Grants No. 11772177, No. 11704383, and No. 11547194; the West Light Foundation of the Chinese Academy of Sciences under Grant No. XAB2016B73; the Applied Basic Research Foundation of Shanxi Province under Grand No. 201701D121011.


\begin{thebibliography}{99}
\bibitem{NRCooper} N. R. Cooper, Rapidly rotating atomic gases, Adv. Phys. \textbf{57}, 539 (2008).
\bibitem{ALFetter} A. L. Fetter, Rotating trapped Bose-Einstein condensates, Rev. Mod. Phys. \textbf{81}, 647 (2009).
\bibitem{JDalibard} K. W. Madison, F. Chevy, W. Wohlleben, and J. Dalibard, Vortex formation in a stirred Bose-Einstein condensate, Phys. Rev. Lett. \textbf{84}, 806 (2000).
\bibitem{WKetterle} J. R. Abo-Shaeer, C. Raman, J. M. Vogels, and W. Ketterle. Observation of vortex lattices in Bose-Einstein condensates. Science, \textbf{292}, 476 (2001).
\bibitem{WKetterle2} M. W. Zwierlein, J. R. Abo-Shaeer, A. Schirotzek, C. H. Schunck, and W. Ketterle, Vortices and superfluidity in a strongly interacting Fermi gas, Nature \textbf{435}, 1047 (2005).
\bibitem{JWPan} X.-C. Yao, H.-Z. Chen, Y.-P. Wu, X.-P. Liu, X.-Q. Wang, X. Jiang, Y. Deng, Y.-A. Chen, and J.-W. Pan. Observation of coupled vortex lattices in a mass-imbalance Bose and Fermi superfluid mixture, Phys. Rev. Lett. \textbf{117}, 145301 (2016).
\bibitem{EACornell} V. Schweikhard, I. Coddington, P. Engels, S. Tung, and E. A. Cornell, Vortex-lattice dynamics in rotating spinor Bose-Einstein condensates, Phys. Rev. Lett. \textbf{93}, 210403 (2004).
\bibitem{MUeda} K. Kasamatsu, M. Tsubota, and M. Ueda, Spin textures in rotating two-component Bose-Einstein condensates, Phys. Rev. A \textbf{71}, 043611 (2005).
\bibitem{NKWilkin} N. K. Wilkin and J. M. F. Gunn, Condensation of ¡°composite bosons¡± in a rotating BEC, Phys. Rev. Lett. \textbf{84}, 6 (2000).
\bibitem{TLHo} T.-L. Ho, Bose-Einstein condensates with large number of vortices, Phys. Rev. Lett. \textbf{87}, 060403 (2001).
\bibitem{PZoller} B. Paredes, P. Fedichev, J. I. Cirac, and P. Zoller. $\frac{1}{2}$-Anyons in small atomic Bose-Einstein condensates, Phys. Rev. Lett. \textbf{87}, 010402 (2001).
\bibitem{EACornell2} V. Schweikhard, I. Coddington, P. Engels, V. P. Mogendorff, and E. A. Cornell. Rapidly rotating Bose-Einstein condensates in and near the lowest Landau level, Phys. Rev. Lett. \textbf{92}, 040404 (2004).
\bibitem{RJDonnelly} R. J. Donnelly, \emph{Quantized Vortices in Helium II} (Cambridge University Press, 2005).
\bibitem{VMVinokur} G. Blatter, M. V. Feigel'man, V. B. Geshkenbein, A. I. Larkin, and V. M. Vinokur, Vortices in high-temperature superconductors, Rev. Mod. Phys. 66, 1125 (1994).
\bibitem{MRho} G. E. Brown and M. Rho, \emph{The multifaceted skyrmion} (World Scientific, Singapore, 2010).
\bibitem{YTokura} N. Nagaosa and Y. Tokura, Topological properties and dynamics of magnetic skyrmions, Nat. Nano. \textbf{8}, 899 (2013).
\bibitem{SMGirvin} R. E. Prange and S. M. Girvin (eds.), \emph{The Quantum Hall Effect}, 2nd ed. (Springer, Berlin, 1990).
\bibitem{APinczuk} S. D. Sarma and A. Pinczuk (eds.), \emph{Perspectives in Quantum Hall Effects: Novel Quantum Liquids in Low-Dimensional Semiconductor Structures} (Wiley, New York, 1997).
\bibitem{IBSpielman} Y.-J. Lin, R. L. Compton, K. Jim\'{e}nez-Garc\'{\i}a, J. V. Porto, and I. B. Spielman, Synthetic magnetic fields for ultracold neutral atoms, Nature \textbf{462}, 628 (2009).
\bibitem{IBSpielman2} Y.-J. Lin, K. Jim\'{e}nez-Garc\'{\i}a, and I. B. Spielman, Spin-orbit-coupled Bose-Einstein condensates, Nature \textbf{471}, 83 (2011).
\bibitem{Jing-Zhang} P. Wang, Z.-Q. Yu, Z. Fu, J. Miao, L. Huang, S. Chai, H. Zhai, and J. Zhang, Spin-orbit coupled degenerate Fermi gases, Phys. Rev. Lett. \textbf{109}, 095301 (2012).
\bibitem{MWZwierlein} L. W. Cheuk, A. T. Sommer, Z. Hadzibabic, T. Yefsah, W. S. Bakr, and M. W. Zwierlein, Spin-injection spectroscopy of a spin-orbit coupled Fermi gas, Phys. Rev. Lett. \textbf{109}, 095302 (2012).
\bibitem{Jing-Zhang2} L. Huang, Z. Meng, P. Wang, P. Peng, S.-L. Zhang, L. Chen, D. Li, Q. Zhou, and J. Zhang, Experimental realization of two-dimensional synthetic spin-orbit coupling in ultracold Fermi gases, Nat. Phys. \textbf{12}, 540 (2016).
\bibitem{Shuai-Chen} Z. Wu, L. Zhang, W. Sun, X.-T. Xu, B.-Z. Wang, S.-C. Ji, Y. Deng, S. Chen, X.-J. Liu, and J.-W. Pan, Realization of two-dimensional spin-orbit coupling for Bose-Einstein condensates, Science \textbf{354}, 83 (2016).
\bibitem{Jun-Ye} S. Kolkowitz, S. L. Bromley, T. Bothwell, M. L. Wall, G. E. Marti, A. P. Koller, X. Zhang, A. M. Rey, and J. Ye, Spin-orbit-coupled fermions in an optical lattice clock, Nature \textbf{542}, 66 (2017).
\bibitem{WKetterle6} J.-R. Li, J. Lee, W. Huang, S. Burchesky, B. Shteynas, F. \c{C}. Top, A. O. Jamison, and W. Ketterle, A stripe phase with supersolid properties in spin-orbit-coupled Bose-Einstein condensates, Nature \textbf{543}, 91 (2017).
\bibitem{VGalitski} V. Galitski and I. B. Spielman, Spin-orbit coupling in quantum gases, Nature \textbf{494}, 49 (2013).
\bibitem{JDalibard2} J. Dalibard, F. Gerbier, G. Juzeli\={u}nas, and P. \"{O}hberg, Colloquium: artificial gauge potentials for neutral atoms, Rev. Mod. Phys. \textbf{83}, 1523 (2011).
\bibitem{NGoldman} N. Goldman, G. Juzeli\={u}nas, P. \"{O}hberg, and I. B. Spielman, Light-induced gauge fields for ultracold atoms, Rep. Prog. Phys. \textbf{77}, 126401 (2014).
\bibitem{Hui-Zhai} H. Zhai, Spin-orbit coupled quantum gases, Int. J. Mod. Phys. B \textbf{26}, 1230001 (2012).
\bibitem{Hui-Zhai2} H. Zhai, Degenerate quantum gases with spin-orbit coupling: a review, Rep. Prog. Phys. \textbf{78}, 026001 (2015).
\bibitem{WYi} W. Yi, W. Zhang, and X. L. Cui, Pairing superfluidity in spin-orbit coupled ultracold Fermi gases, Sci. China: Phys. Mech. Astron. \textbf{58}, 014201 (2015).
\bibitem{Jing-Zhang3} J. Zhang, H. Hu, X. J. Liu, and H. Pu, Fermi gases with synthetic spin-orbit coupling, Annu. Rev. Cold At. Mol. \textbf{2}, 81 (2014).
\bibitem{Congjun-Wu} C. Wu, Unconventional Bose-Einstein condensations beyond the ``no-node" theorem, Mod. Phys. Lett. B \textbf{23}, 1 (2009).
\bibitem{VGalitski2} J. Radi\'{c}, T. A. Sedrakyan, I. B. Spielman, and V. Galitski, Vortices in spin-orbit-coupled Bose-Einstein condensates, Phys. Rev. A \textbf{84}, 063604 (2011).
\bibitem{ALFetter2} A. L. Fetter, Vortex dynamics in spin-orbit-coupled Bose-Einstein condensates, Phys. Rev. A \textbf{89}, 023629 (2014).
\bibitem{JHHan} X.-Q. Xu and J. H. Han, Spin-orbit coupled Bose-Einstein condensate under rotation. Phys. Rev. Lett. 107, 200401 (2011).
\bibitem{Congjun-Wu2} X.-F. Zhou, J. Zhou, and C. Wu. Vortex structures of rotating spin-orbit-coupled Bose-Einstein condensates. Phys. Rev. A 84, 063624 (2011).
\bibitem{IBSpielman3} I. B. Spielman, Raman processes and effective gauge potentials, Phys. Rev. A \textbf{79}, 063613 (2009).
\bibitem{SStringari2} C. Qu and S. Stringari, Angular Momentum of a Bose-Einstein Condensate in a Synthetic Rotational Field, Phys. Rev. Lett. \textbf{120}, 183202 (2018).
\bibitem{SStringari} S. Stringari, Diffused vorticity and moment of inertia of a spin-orbit coupled Bose-Einstein condensate, Phys. Rev. Lett. \textbf{118}, 145302 (2017).
\bibitem{AAAbrikosov} A. A. Abrikosov, Zh. Eksp. Teor. Fiz. Sov. Phys. JETP, \textbf{5}, 1174 (1957).
\bibitem{YAharonov}Y. Aharonov and D. Bohm. Significance of electromagnetic potentials in the quantum theory. Phys. Rev. \textbf{115}, 485 (1959).
\bibitem{IBSpielman4}Y.-J. Lin and I. B. Spielman. Synthetic gauge potentials for ultracold neutral atoms. J. Phys. B: At. Mol. Opt. Phys. \textbf{49}, 183001 (2016).
\bibitem{MSIoffe} Y. V. Gott, M. S. Ioffe, and V. G. Telkovskii, Some new results on confinement in magnetic traps, Nucl. Fusion Suppl. \textbf{3}, 1045 (1962).
\bibitem{DEPritchard} D. E. Pritchard, Phys. Rev. Lett. \textbf{51}, 1336 (1983).
\bibitem{WKetterle3} M.-O. Mewes, M. R. Andrews, N. J. van Druten, D. M. Kurn, D. S. Durfee, and W. Ketterle, Bose-Einstein condensation in a tightly confining dc magnetic trap, Phys. Rev. Lett. \textbf{77}, 416 (1996).
\bibitem{WKetterle4} A. E. Leanhardt, A. G\"{o}rlitz, A. P. Chikkatur, D. Kielpinski, Y. Shin, D. E. Pritchard, and W. Ketterle, Imprinting vortices in a Bose-Einstein condensate using topological phases, Phys. Rev. Lett. \textbf{89}, 190403 (2002).
\bibitem{WKetterle5} A. E. Leanhardt, Y. Shin, D. Kielpinski, D. E. Pritchard, and W. Ketterle, Coreless vortex formation in a spinor Bose-Einstein condensate, Phys. Rev. Lett. \textbf{90}, 140403 (2003).
\bibitem{MMottonen} M. M\"{o}tt\"{o}nen, V. Pietil\"{a}, and S. M. M. Virtanen, Vortex pump for dilute Bose-Einstein condensates, Phys. Rev. Lett. \textbf{99}, 250406 (2007).
\bibitem{GJuzeliunas} D. L. Campbell, G. Juzeli\={u}nas, and I. B. Spielman, Realistic Rashba and Dresselhaus spin-orbit coupling for neutral atoms. Phys. Rev. A \textbf{84}, 025602 (2011).
\bibitem{ZFXu} Z. F. Xu and L. You, Dynamical generation of arbitrary spin-orbit couplings for neutral atoms, Phys. Rev. A \textbf{85}, 043605 (2012).
\bibitem{LYou} Z.-F. Xu, L. You, and M. Ueda, Atomic spin-orbit coupling synthesized with magnetic-field-gradient pulses, Phys. Rev. A \textbf{87}, 063634 (2013).
\bibitem{GJuzeliunas2} B. M. Anderson, I. B. Spielman, and G. Juzeli\={u}nas, Magnetically generated spin-orbit coupling for ultracold atoms, Phys. Rev. Lett. \textbf{111}, 125301 (2013).
\bibitem{Ruquan-Wang} X. Luo, L. Wu, J. Chen, Q. Guan, K. Gao, Z.-F. Xu, L. You, and R. Wang, Tunable atomic spin-orbit coupling synthesized with a modulating gradient magnetic field, Sci. Rep. \textbf{6}, 18983 (2016).
\bibitem{JDalibard3} T. Yefsah, R. Desbuquois, L. Chomaz, K. J. G\"{u}nter, and J. Dalibard, Exploring the thermodynamics of a two-dimensional Bose gas, Phys. Rev. Lett. \textbf{107}, 130401 (2011).
\bibitem{JHHan2} X.-Q. Xu and J. H. Han, Emergence of chiral magnetism in spinor Bose-Einstein condensates with Rashba coupling, Phys. Rev. Lett. \textbf{108}, 185301 (2012).
\bibitem{GBaym} T. Ozawa and G. Baym, Striped states in weakly trapped ultracold Bose gases with Rashba spin-orbit coupling, Phys. Rev. A \textbf{85}, 063623 (2012).
\bibitem{XFZhang} X.-F. Zhang, M. Kato, W. Han, S.-G. Zhang, and H. Saito. Spin-orbit-coupled Bose-Einstein condensates held under a toroidal trap. Phys. Rev. A 95, 033620 (2017).
\bibitem{Yongping-Zhang} A. C. White, Y. Zhang, and T. Busch. Odd-petal-number states and persistent flows in spin-orbit-coupled Bose-Einstein condensates. Phys. Rev. A 95, 041604(R) (2017).
\bibitem{TLHo2} E. J. Mueller and T.-L. Ho, Two-component Bose-Einstein condensates with a large number of vortices, Phys. Rev. Lett. \textbf{88}, 180403 (2002).
\bibitem{MUeda2} K. Kasamatsu, M. Tsubota, and M. Ueda, Vortex phase diagram in rotating two-component Bose-Einstein condensates, Phys. Rev. Lett. \textbf{91}, 150406 (2003).
\bibitem{NRCooper2} N. R. Cooper, E. H. Rezayi, and S. H. Simon, Vortex lattices in rotating atomic Bose gases with dipolar interactions, Phys. Rev. Lett. \textbf{95}, 200402 (2005).
\bibitem{Hui-Zhai3} J. Zhang and H. Zhai, Vortex lattices in planar Bose-Einstein condensates with dipolar interactions, Phys. Rev. Lett. \textbf{95}, 200403 (2005).
\bibitem{LSantos} S. Sinha, R. Nath, and L. Santos, Trapped two-dimensional condensates with synthetic spin-orbit coupling, Phys. Rev. Lett. \textbf{107}, 270401 (2011).
\bibitem{Hui-Hu} H. Hu, B. Ramachandhran, H. Pu, and X.-J. Liu, Spin-orbit coupled weakly interacting Bose-Einstein condensates in harmonic traps, Phys. Rev. Lett. \textbf{108}, 010402 (2012).
\bibitem{LYou2} Z. F. Xu, R. L\"{u}, and L. You, Emergent patterns in a spin-orbit-coupled spin-2 Bose-Einstein condensate, Phys. Rev. A \textbf{83}, 053602 (2011).
\bibitem{SCGou} S.-W. Su, I.-K. Liu, Y.-C. Tsai, W. M. Liu, and S.-C. Gou, Crystallized half-skyrmions and inverted half-skyrmions in the condensation of spin-1 Bose gases with spin-orbit coupling, Phys. Rev. A \textbf{86}, 023601 (2012).
\bibitem{MUeda3} Z. F. Xu, Y. Kawaguchi, L. You, and M. Ueda, Symmetry classification of spin-orbit-coupled spinor Bose-Einstein condensates, Phys. Rev. A \textbf{86}, 033628 (2012).
\bibitem{Wei-Han} W. Han, X.-F. Zhang, S.-W. Song, H. Saito, W. Zhang, W.-M. Liu, and S.-G. Zhang, Double-quantum spin vortices in SU(3) spin-orbit-coupled Bose gases, Phys. Rev. A \textbf{94}, 033629 (2016).
\bibitem{Congjun-Wu3}X.-F. Zhou, Z.-W. Zhou, C. Wu, and G.-C. Guo. In-plane gradient-magnetic-field-induced vortex lattices in spin-orbit-coupled Bose-Einstein condensations. Phys. Rev. A \textbf{91}, 033603 (2015).
\end{thebibliography}
\end{document}